\documentclass[twocolumn,showpacs,preprintnumbers,amsmath,amssymb]{revtex4}
\input psfig.sty

\usepackage{graphicx}
\usepackage{dcolumn}
\usepackage{bm}

\begin{document}



\title{Brane World Cosmologies and Statistical Properties of
Gravitational Lenses} 
 
\author{D. Jain} \email{deepak@ducos.ernet.in}  \author{A. Dev} \email{abha@ducos.ernet.in} 
\affiliation{%
Department of Physics and Astrophysics, University of Delhi,
Delhi-110007, India  
}%

 
\author{J. S. Alcaniz} \email{alcaniz@astro.washington.edu}
\affiliation{Astronomy Department, University of Washington, Seattle,
Washington, 98195-1580, USA
}%

\date{\today}

\begin{abstract}
Brane world cosmologies seem to provide an alternative explanation for the present accelerated 
stage of the Universe with no need to invoke either a cosmological constant or an exotic 
\emph{quintessence} component. In this paper we investigate statistical properties of gravitational 
lenses for some particular scenarios based on this large scale modification of gravity. We show 
that a large class of such models are compatible with the current lensing data for values of the
matter density 
parameter $\Omega_{\rm{m}} \leq 0.94$ ($1\sigma$). If one fixes $\Omega_{\rm{m}}$ to be $\simeq 
0.3$, as suggested by most of the dynamical estimates of the quantity of matter in the Universe, 
the 
predicted number of lensed quasars requires a slightly open universe with a crossover distance 
between the 4 and 5-dimensional gravities of the order of $1.76 H_o^{-1}$.
\end{abstract}

\maketitle

\section{Introduction}

The results of observational cosmology in the last years have
opened up an unprecedented   
opportunity to test the veracity of a number of cosmological scenarios
as well as to establish a 
more solid connection between particle physics and cosmology. The most
remarkable finding among  
these results comes from distance measurements of type Ia supernovae
(SNe Ia) that suggest that the  
expansion of the Universe is speeding up, not slowing down
\cite{perlmutter}. As widely known  
such a result poses a crucial problem for all CDM models since their
generic prediction is a  
decelerating universe ($q_{o} > 0$), whatever the sign adopted for the
curvature parameter.  
Indirectly, similar results have also been obtained, independent of
the SNe Ia analyses, by  
combining the latest galaxy clustering data with CMB measurements \cite{efs}.
 
To reconcile these observational results with theory, cosmologists
have proposed more general  
models containing a negative-pressure dark component that would be
responsible for the present  
accelerated stage of the Universe. Although a large number of
pieces of observational evidence have  
consistently suggested a universe composed of $\sim 2/3$ of dark 
energy, the exact nature of this new component is not well understood
at present. Among the several  
candidates for dark energy discussed in the recent 
literature, the simplest and most theoretically appealing possibility
is the vacuum energy or  
cosmological constant. Despite the serious problem that arises when
one considers a nonzero vacuum  
energy \cite{wein}, models with a relic cosmological constant
($\Lambda$CDM) seem to be our best  
description of the observed universe, being considered as a serious
candidate for standard cosmology. 

On the other hand, motivated by particle physics considerations, there
has been growing interest in  
cosmological models based on the framework of brane-induced gravity 
\cite{ark,dvali,deff,randall}. The general principle behind such models is that our 
4-dimensional Universe would be a surface or a brane embedded into a
higher dimensional bulk  
space-time on which gravity can propagate. In some of these scenarios, there is a certain crossover 
scale $r_c$ that defines what kind of gravity an observer on the brane will observe. For distances 
shorter than $r_c$, such an observer will measure the usual 4-dimensional
gravitational $1/r^{2}$ force  
whereas for distances larger than $r_c$ the  gravitational force
follows the 5-dimensional  
$1/r^{3}$ behavior. In this way, gravity gets weaker at cosmic
distances and, therefore, it is  
natural to think that such an effect has some implications on the
dynamics of the Universe  
\cite{deff1}. 

Several aspects of brane world cosmologies have been explored in the recent literature. For 
example, the issue related to the cosmological constant problem has
been addressed \cite{ccp} as well as    
evolution of cosmological perturbations in the gauge-invariant
formalism \cite{brand}, cosmological  
phase transitions \cite{cpt}, inflationary solutions \cite{cpt1},
baryogenesis \cite{dvali99},   
stochastic background of gravitational waves \cite{hogan1}, singularity, 
homogeneity, flatness and entropy  
problems \cite{aaa},  among others (see \cite{hogan} for a 
discussion on the different perspectives of brane world models). From
the observational viewpoint,  
however, the present situation is somewhat controversial. While the
authors of Refs. \cite{deffZ,dnew}  
have shown that such models are in agreement with the most recent
cosmological observations (for  
example, they found that a flat universe with $\Omega_{\rm{m}} = 0.3$
and $r_c \simeq 1.4H_o^{-1}$ is consistent with the currently SNe Ia + CMB data), the authors
of Ref. \cite{avelino} have  
claimed that a larger sample of SNe Ia data can also be used to rule out these models
at least at the 2$\sigma$  
level. Recently, one of us \cite{alcaniz} used
measurements of the angular size  
of high-$z$ compact radio sources to show that the best fit model for
these data is a slightly closed universe with $\Omega_{\rm{m}} \simeq 0.06$ and a crossover 
radius of the order of 0.94$H_o^{-1}$.

For the reasons presented earlier, the comparison between any
alternative cosmology and  
$\Lambda$CDM models is very important. In this concern, statistical
properties of gravitational  
lenses may be an interesting tool because, as is well known, they
provide restrictive limits on  
the vaccum energy contribution (see, for instance, \cite{1CSK}). On the
other hand, in brane world  
models the distance to an object at a given redshift $z$ is smaller
than the distance to the same  
object in $\Lambda$CDM models (assuming the 
same value of $\Omega_{\rm{m}}$). Therefore, we expect that
the constraints coming from lensing  
statistics will be weaker for these models than for their $\Lambda$CDM counterparts.

In this paper, we explore the implications of gravitationally lensed
QSOs for models based on the  
framework of the brane-induced gravity of Dvali {\it et al.}
\cite{dvali} that have been recently  
proposed in Refs. \cite{deff,deff1}. We restrict our analysis to
accelerated models, or  
equivalently, models that have a ``self-inflationary" solution with $H
\sim r_c^{-1}$ ($H$ is the  
Hubble parameter). As explained in \cite{deff}, in such scenarios, the bulk gravity sees its own
curvature term on the brane as a negative-pressure dark component and accelerates the Universe.

This paper is organized as follows. In section 2, we present the basic field equations and distance 
formulas relevant for our analysis. We then proceed to analyze the constraints from lensing  
statistics on these models in section 3. In section 4 our main conclusions are presented. 

\section{The Model: Basic equations and distance Formulas}

The Friedmann's equation 
for the kind of models we are considering is \cite{deff1,deffZ}
\begin{equation} 
\left[\sqrt{\frac{\rho}{3M_{pl}^{2}} + \frac{1}{4r_{c}^{2}}} +
\frac{1}{2r_{c}}\right]^{2} = H^{2}  
+ \frac{k}{R(t)^{2}},
\end{equation} 
where $\rho$ is the energy density of the cosmic fluid, $k = 0, \pm 1$
is the spatial curvature,   
$M_{pl}$ is the Planck mass and $r_c = M_{pl}^{2}/2M_{5}^{3}$ is the
crossover scale defining the  
gravitational interaction among particles located on the brane ($M_5$
is the 5-dimensional reduced  
Planck mass). From the above equation we find that the normalization
condition is given by 
\begin{equation}
\Omega_k + \left[\sqrt{\Omega_{\rm{r_c}}} + \sqrt{\Omega_{\rm{r_c}} +
\Omega_{\rm{m}}}\right]^{2} =  
1
\end{equation}
where $\Omega_{\rm{m}}$ and $\Omega_k$ are, respectively, the matter
and curvature density  
parameters (defined in the usual way) and 
\begin{equation}
\Omega_{\rm{r_c}} = 1/4r_c^{2}H^{2},
\end{equation} 
is the density parameter associated with the crossover radius $r_c$. For
a flat universe, the  
normalization condition becomes \cite{deffZ}
\begin{equation}
\Omega_{\rm{r_c}} = \left(1 - \Omega_{\rm{m}} \over 2 \right)^{2}
\quad \mbox{for} \quad   
\Omega_{\rm{r_c}} < 1 \quad \mbox{and} \quad \Omega_{\rm{m}} < 1.
\end{equation}

In order to derive the constraints from lensing statistics in the next section
we shall use the concept of  
angular diameter distance, $D_A(z)$. Such a quantity can be easily
obtained in the following way:  
consider that photons are emitted by a source with coordinate $r =
r_1$ at time $t_1$ and are  
received at time $t_o$ by an observer located at coordinate $r =
0$. The emitted radiation will  
follow null geodesics on which the dimensionless comoving coordinates
$\theta$ and $\phi$ are  
constant. The comoving distance of the source is defined by
\begin{equation}
r_1 = \int_{t_1}^{t_o} {dt \over R(t)} = \int_{R(t)}^{R_o} {dR \over
\dot{R}(t)R(t)}. 
\end{equation}

From Eqs. (1) and (5), it is possible to show that the comoving
distance $r_1(z)$ can be written as \cite{alcaniz}
\begin{equation}  
r_1(z) = \frac{1}{R_o H_o |\Omega_k|^{1/2}}\sum\left[|\Omega_k|^{1/2}
\int_{x'}^{1} {dx \over  
x^{2}f(\Omega_{j}, x)}\right], 
\end{equation} 
where the subscript $o$ denotes present day quantities, $x' = {R(t)
\over R_o} = (1 + z)^{-1}$ is a  
convenient integration variable and the function $\sum(r)$ is defined
by one of the following  
forms: $\sum(r) = \mbox{sinh}(r)$, $r$, and $\mbox{sin}(r)$,
respectively, for open, flat and  
closed geometries. The dimensionless function $f(\Omega_{j}, x)$ is given by 
\begin{equation}
f(\Omega_{j}, x) = \left[\Omega_k x^{-2} + \left(\sqrt{\Omega_{\rm{r_c}}} + 
\sqrt{\Omega_{\rm{r_c}} 
+ \Omega_{\rm{m}}x^{-3}}\right)^{2}\right]^{1/2},
\end{equation}
where $j$ stands for $m$, $r_c$ and $k$. 

The angular diameter distance to a light source at $r = r_1$ and $t =
t_1$ and observed at $r = 0$  
and $t = t_o$ is defined as the ratio of the source diameter to its
angular diameter, i.e., 
\begin{equation}
D_{A} = {\ell \over \theta} =  R(t_1)r_1 .
\end{equation}
In the general case, the angular diameter distance, $D_{LS}(z_L, z_S) =
{R_or(z_L,  
z_S) \over (1 + z_S)}$, between two objects, for example, a lens at $z_L$ and a
source (galaxy) at $z_S$, reads  
\begin{eqnarray}
D_{LS}(z_L, z_S) & = & \frac{H_o^{-1}}{(1 +
z_S)|\Omega_k|^{1/2}} \times \\ \nonumber & & \times  
\sum \left[|\Omega_k|^{1/2} \int_{x'_S}^{x'_L} {dx \over 
x^{2} f(\Omega_{j}, x)}\right] .
\end{eqnarray}

\section{Constraints from Lensing Statistics}
 
In this paper we work with a sample of 867 ($z > 1$) high luminosity 
optical quasars which include 5 lensed quasars. These data are taken 
from optical lens surveys such as the
HST Snapshot survey \cite{HST}, the Crampton survey \cite{Crampton}, 
the Yee survey \cite{Yee}, Surdej survey \cite{Surdej}, the NOT Survey
\cite{Jaunsen} and the FKS survey \cite{FKS} .
Since the lens surveys and quasar catalogs usually use V
magnitudes, we transform $m_{V}$ to B-band magnitude by using $B-V =
0.2$  as suggested by Bahcall {\it et al.} \cite{Bahcall}.

The differential probability $d\tau$ of a beam having a lensing 
event in traversing $dz_{L}$ is \cite{TOG,FFKT}
\begin{equation}
d\tau = F^{*}(1 +
z_{L})^{3}\left({D_{OL}D_{LS}\over R_0 D_{OS}}\right)^{2}
{1\over R_0}{dt \over dz_{L}} dz_{L},
\end{equation}
where 
\begin{equation} {cdt\over dz_L} = {H_o^{-1} \over (1 + z_L) f(\Omega_j, x_L)},
\end{equation}
and
\begin{equation}
F^* = {16\pi^{3}\over{c
H_{0}^{3}}}\phi_\ast v_\ast^{4}\Gamma\left(\alpha + {4\over\gamma} +1\right).
\end{equation}
$D_{OL}$, $D_{OS}$ and $D_{LS}$ are, respectively, the angular diameter distances from the 
observer to the lens, from the observer to the source and between the lens and the source. For 
simplicity we use the Singular Isothermal Model (SIS) for the lens mass distribution.
The Schechter luminosity function is adopted and lens parameters for E/SO galaxies are taken from 
Loveday {\it et al.} \cite{loveday} (LPEM), i.e., $\phi_{\ast} = 3.2\pm 0.17 h^{3}\,10^{-3} 
{\rm{Mpc}}^{-3}$, $\alpha = 0.2$, $\gamma = 4$, $v_{\ast} = 205.3\,{\rm{km/s}}$ and
$F^{*} = 0.010$. It is worth mentioning that, although the recent galaxy surveys  have increased 
considerably our knowledge of the galaxy luminosity function, they do not classify the galaxies by 
their morphological type \cite{chris}. In this work we restrict ourselves to the LPEM 
parameters because they have been derived in a highly correlated manner and they also take into
account the morphological distribution of the E/S0 galaxies {\cite{c}}.

The differential optical depth of lensing in traversing
$dz_{L}$ with angular separation between $\phi$ and $\phi + d\phi$, is
given by
\begin{eqnarray}
\frac{d^{2}\tau}{dz_{L}d\phi}d\phi dz_{L} 
&=& F^{*}\,(1 + z_{L})^{3}\,\left({{D_{OL} D_{LS}}\over{ R_{o} 
D_{OS}}}\right)^{2}\,\frac{1}{R_{o}}\, \frac{dt}{dz_{L}} \nonumber \\
&& \times \, \frac{\gamma/2}{\Gamma(\alpha+1+\frac{4}{\gamma})} 
\left(\frac{D_{OS}}{D_{LS}}\phi\right)^{\frac{\gamma}{2}(\alpha+1+\frac{4}{\gamma})} \nonumber \\
&& \times \,{\rm{exp}}\left[-\left(\frac{D_{OS}}{D_{LS}}\phi\right)^{\frac{\gamma}{2}}\right] 
\frac{d\phi}{\phi} 
dz_{L}
\label{diff}
\end{eqnarray}
where $\phi = \Delta\theta/8\pi (v_{\ast}/c)^{2}$, with the velocity dispersion $v_{\ast}$ 
corresponding to the characteristic luminosity $L_{\ast}$ in the Schechter luminosity function. The 
total optical depth is obtained by integrating $d\tau$ along the
line of sight from $z = 0$ ($z_O$) to $z_S$. One obtains
\begin{equation}
\tau(z_S) = \frac{F^{*}}{30} \left[D_{OS}(1+z_{L})\right]^{3}\, R_{o}^{3}.
\label{atau}
\end{equation}
Figure 1 shows the normalized optical depth as a function of the source redshift ($z_S$) for 
$\Omega_{\rm{m}} = 0.3$ and 
values of $\Omega_{\rm{r_c}} =$ 0.1, 0.2 and 0.3. For comparison, the standard prediction 
($\Omega_{\rm{r_c}} = 0$) is also displayed.
Note that, at higher redshifts ($z > 2.5$), an increase in 
$\Omega_{\rm{r_c}}$ at fixed $\Omega_{\rm{m}}$ tends to reduce the optical depth for lensing. For 
example, at $z_S = 3.0$ the value of $\tau/F^{*}$ for $\Omega_{\rm{r_c}} = 0.1$ is down from the 
standard value by a factor of $\sim$ 1.10. This 
decrease of the optical depth as the value of $\Omega_{\rm{r_c}}$ is increased (at a fixed 
$z_S$ 
and $\Omega_{\rm{m}}$) occurs because, at high redshift, say $z > 2.5$, the distance between two 
redshifts (e.g., 
$z_O$ and $z_S$) is smaller for higher values of $\Omega_{\rm{r_c}}$.

\begin{figure}
\centerline{\psfig{figure=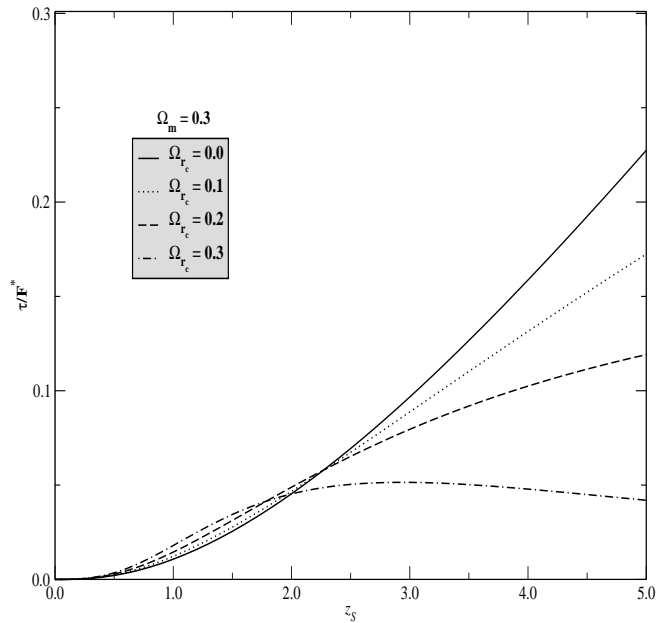,width=3.5truein,height=3.5truein,angle=-90}
\hskip 0.1in} 
\caption{The normalized optical depth ($\tau/F^{*}$) as a function of the source redshift ($z_S$) 
for some selected  values of $\Omega_{\rm{r_c}}$. In all curves, the value of the matter density 
parameter has been fixed ($\Omega_{\rm{m}} = 0.3$).}
\end{figure}

In order to obtain the correct lensing probability we have made two corrections to the optical 
depth, 
namely, magnification bias and selection function. Magnification bias, ${\bf B}(m,z)$, is 
considered in order to take into account the increase in the
apparent brightness of a quasar due to lensing which, in turn, increases  the
expected number of lenses in flux limited sample.
The bias factor for a quasar at redshift $z$ with apparent magnitude
$m$ is  given by \cite{FFKT,1CSK}
\begin{equation}
{\bf B}(m,z) = M_{0}^{2}\, \emph{B}(m,z,M_{0},M_{2}),
\label{bias}
\end{equation} 
where
\begin{eqnarray}
\emph{B}(m,z,M_{1},M_{2})& =&
2\,\left(\frac{dN_{Q}}{dm}\right)^{-1}\int_{M_{1}}^{M_{2}}\frac{dM}{M^{3}}\, \\
&& \times \,\frac{dN_{Q}}{dm}(m+2.5\log(M),z). \nonumber
\label{bias1}
\end{eqnarray}
In the above equation $({dN_{Q}(m,z)}/{dm})$ is the measure of number of quasars with magnitudes 
in the interval $(m,m+dm)$ at redshift $z$. Since we are modeling the lens by a SIS profile, $M_0 = 
2$, we adopt $M_{2} = 10^{4}$ in the numerical computation.

We use Kochanek's ``best model'' \cite{1CSK} for the quasar luminosity
function:
\begin{equation}
\frac{dN_{Q}}{dm}(m,z) \propto
(10^{-a(m-\overline{m})}+10^{-b(m-\overline{m})})^{-1}
\label{lum},
\end{equation}
where
\begin{equation}
\overline{m} = \left\{ \begin{array}{ll}
                   m_{o}+(z-1)    & \mbox{for $z < 1$} \\
                   m_{o}          & \mbox{for $1 < z \leq 3$} \\
                   m_{o}-0.7(z-3) & \mbox{for $z > 3$}
                   \end{array}
               \right. \
\end{equation}
and we assume $a = 1.07 \pm 0.07$, $b = 0.27 \pm 0.07$ and $m_{o} = 18.92 \pm 0.16$ at B magnitude 
\cite{1CSK}.

The magnitude corrected probability, $p_{i}$, for a given quasar $i$ at $z_{i}$ and 
apparent magnitude $m_{i}$ to be lensed is
\begin{equation}
p_{i} = \tau(z_{i}){\bf B}(m_{i},z_{i}).
\label{prob1}
\end{equation}

Due to selection effects the survey can only detect lenses with magnifications larger than a 
certain magnitude $M_{f}$. It can be shown that the corrected lensing probability and image 
separation distribution function for a single source at redshift $z_{S}$ are \cite{1CSK,2CSK}
\begin{equation}
p^{'}_{i}(m,z) = p_{i}\int \frac{ d(\Delta\theta)\,
p_{c}(\Delta\theta)\emph{B}(m,z,M_{f}(\Delta\theta),M_{2})}{\emph{B}(m,z,M_{0},M_{2})}
\label{prob2}
\end{equation}
and
\begin{equation}
p^{'}_{ci} =
p_{ci}(\Delta\theta)\,\frac{p_{i}}{p_{i}^{'}}\,\frac{\emph{B}(m,z,M_{f}(\Delta
\theta),M_{2})} 
{\emph{B}(m,z,M_{0},M_{2})},
\label{confi}
\end{equation}
where
\begin{equation}
p_{c}(\Delta\theta) =
\frac{1}{\tau(z_{S})}\,\int_{0}^{z_{S}}\frac{d^{2}\tau}{dz_{L}d(\Delta\theta)}
\,dz_{L}
\label{pcphi}
\end{equation}
and 
\begin{equation}
M_{f} = M_{0}(f+1)/(f-1)\quad \mbox{with}\quad  f = 10^{0.4\,\Delta m(\theta)}.
\end{equation}
Equation (\ref{confi}) defines the configuration probability, i.e., the probability that the lensed 
quasar $i$ is lensed with the observed image separation. To obtain selection function corrected 
probabilities, we follow \cite{1CSK} and divide our sample into two parts, namely, the ground based 
surveys and the HST survey.

\begin{figure}
\centerline{\psfig{figure=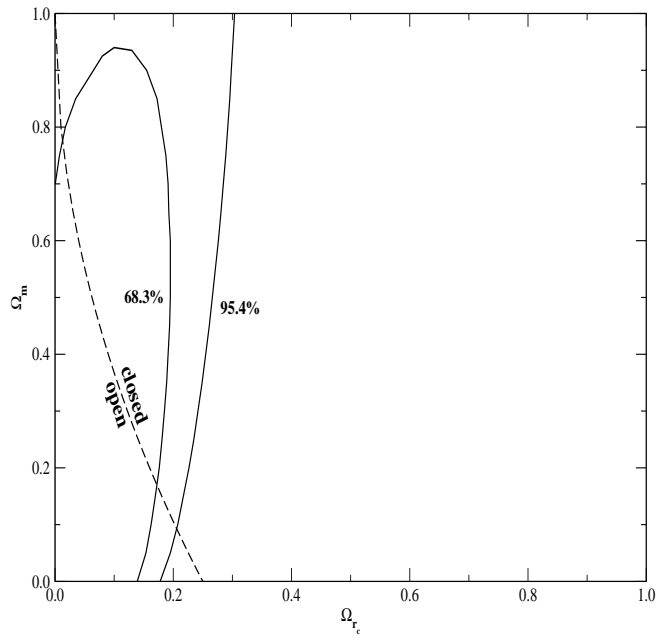,width=3.5truein,height=3.5truein,angle=-90}
\hskip 0.1in} 
\caption{Confidence regions in the plane $\Omega_{\rm{m}} - \Omega_{r_{c}}$ arising from lensing 
statistics. Solid lines indicate contours of constant likelihood at $68\%$ and 
$95.4\%$.}
\end{figure}

In order to constrain the parameters $\Omega_{\rm{m}}$ and $\Omega_{r_{c}}$ we perform a 
maximum-likelihood analysis with the likelihood function given by \cite{1CSK}
\begin{equation}
{\cal{L}} = \prod_{i=1}^{N_{U}}(1-p^{'}_{i})\,\prod_{k=1}^{N_{L}}
p_{k}^{'}\,p_{ck}^{'},
\label{LLF}
\end{equation}
where $N_{L}$ is the number of multiple-imaged lensed quasars, $N_{U}$ is the number of unlensed 
quasars, and $p_{k}^{'}$ and $p_{ck}^{i}$ are the probability of quasar $k$ to be lensed and the 
configuration probability defined, respectively, by Eqs. (\ref{prob2}) and (\ref{confi}). 

Figure 2 shows contours of constant likelihood ($68\%$ and $95.4\%$) in the parameter space 
$\Omega_{r_{c}} - \Omega_{\rm{m}}$. The maximum value of the likelihood function is located at 
$\Omega_{\rm{m}} = 0$ and $\Omega_{r_{c}} = 0.03$. At the 1$\sigma$ level, our analysis requires 
$\Omega_{\rm{m}} 
\leq 0.94$ and $\Omega_{r_{c}} \leq 0.19$. Such a result  
means that a large class of these particular scenarios of brane world cosmology studied here are
compatible with the 
current gravitational lensing data at this confidence level. In Fig. 3a the expected number of 
lensed quasars, $n_L = \sum\, p_{i}^{'}$ (the summation is over a given quasar sample), is 
displayed as a function of $\Omega_{r_{c}}$ with the matter density parameter fixed at 
$\Omega_{\rm{m}} = 0.3$ (as indicated by clustering estimates \cite{calb}). The horizontal dashed
line indicates $n_L = 5$, that is the 
number of lensed quasars 
in our sample.  By this analysis, one finds $\Omega_{r_{c}} \simeq 0.08$, a value that is very 
close to 
that obtained by Deffayet {\it et al.} \cite{deffZ} ($\Omega_{r_{c}} = 0.12$) using SNe Ia and 
CMB data and also with the same fixed value for the matter density parameter. In Fig. 3b we show 
the contour for five lensed quasars in the parametric space $\Omega_{\rm{m}} - \Omega_{r_{c}}$. The 
shadowed horizontal region corresponds to the observed range $\Omega_{\rm{m}} = 0.3 \pm 0.1$ 
\cite{calb}. We observe that the higher the value of $\Omega_{\rm{m}}$ the higher the contribution 
of $\Omega_{r_{c}}$ that is required to fit these data.

\begin{figure}
\centerline{\psfig{figure=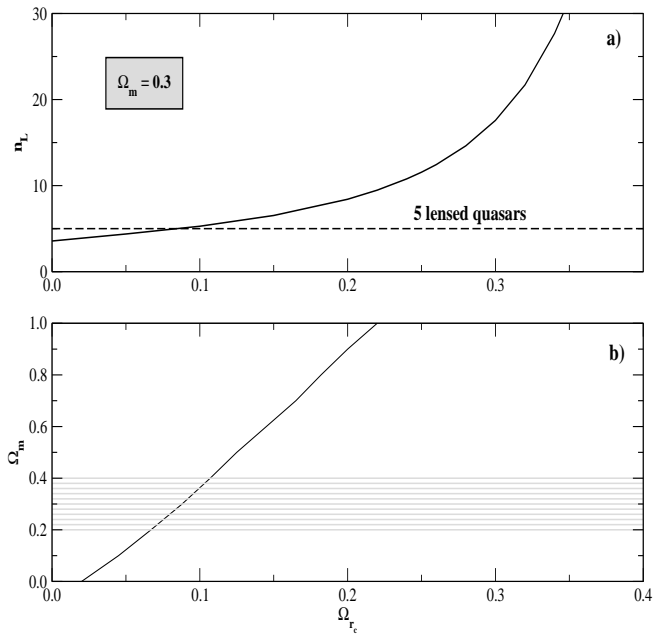,width=3.5truein,height=3.5truein,angle=-90}
\hskip 0.1in} 
\caption{{\bf{a)}} Predicted number of lensed quasars as a function of 
$\Omega_{r_{c}}$ for a fixed value of the matter density parameter ($\Omega_{\rm{m}} = 0.3$) and 
image separation $\Delta \theta \leq 4$. {\bf{b)}} Contour for five lensed quasars in the 
parametric space $\Omega_{\rm{m}} - \Omega_{r_{c}}$. The shadowed horizontal region corresponds to 
the observed range $\Omega_{\rm{m}} = 0.3 \pm 0.1$ \cite{calb}.}
\end{figure}

At this point it is interesting to estimate the value of $r_c$ (the crossover distance between 
4-dimensional and 5-dimensional gravities) from our estimates of $\Omega_{r_{c}}$. In this case, an 
elementary combination of our best fit ($\Omega_{r_{c}} = 0.03$) with Eq. (3) provides
\begin{equation}
r_c \simeq 2.8 H_o^{-1},
\end{equation}
while at the $1\sigma$ level ($\Omega_{r_{c}} \leq 0.19$), we have
\begin{equation}
r_c \geq 1.14 H_o^{-1}.
\end{equation}
The former value is considerably larger than that found in Refs. \cite{deffZ, alcaniz}, i.e., $r_c
\simeq 1.4 
H_o^{-1}$ and $r_c \simeq 0.94 H_o^{-1}$ in analyses involving SNe Ia + CMB and angular size of 
high-$z$ sources data, respectively. However, it is worth mentioning that the estimate of $r_c$ 
obtained in 
Ref. \cite{deffZ} refers to a flat model in which the value of the matter density parameter was 
fixed in $\Omega_{\rm{m}} = 0.3$. As we have seen, by fixing this value for 
$\Omega_{\rm{m}}$, the predicted number 
of lensing quasars (Fig. 3a) requires $\Omega_{r_{c}} \simeq 0.08$ which, in turn, implies $r_c 
\simeq 1.76 H_o^{-1}$.

\section{Conclusion}

The recent observational evidences for a presently accelerated stage of the Universe have 
stimulated 
renewed interest for alternative cosmologies. In general, such models contain an unkown 
negative-pressure dark component that explains the SNe Ia results and reconciles the inflationary 
flatness prediction ($\Omega_{\rm{T}} = 1$) with the dynamical estimates of the quantity of matter 
in the Universe ($\Omega_{\rm{m}} \simeq 0.3 \pm 0.1$). In this paper we have focused our attention 
on another dark energy candidate, one arising from gravitational \emph{leakage} into extra 
dimensions \cite{deff,deff1}. We have shown that some particular scenarios based on this large 
scale modification of gravity are in agreement with the current gravitational lensing data for 
values of $\Omega_{\rm{m}} \leq 0.93$ ($1\sigma$). If one fixes $\Omega_{\rm{m}}$ to be $\simeq
0.3$, the 
predicted number of lensed quasars requires $\Omega_{r_{c}} \simeq 0.08$. This is a slightly open 
universe with a crossover radius of the order of $r_c \simeq 1.76 H_o^{-1}$.

\begin{acknowledgments}
The authors are very grateful to Lee Homer and Carlos J. A. P. Martins for helpful discussions and a 
critical 
reading of the manuscript. JSA is supported by the Conselho Nacional de Desenvolvimento 
Cient\'{\i}fico e Tecnol\'{o}gico (CNPq - Brasil) and CNPq (62.0053/01-1-PADCT III/Milenio).
\end{acknowledgments}


\end{document}